\documentclass[aps,pra]{revtex4}
\usepackage{graphicx}
\usepackage{amsmath}
\usepackage{amsfonts}
\usepackage{amssymb}%
\usepackage{wasysym}

\newcommand{\beq}{\begin{equation}}
\newcommand{\eeq}{\end{equation}}
\newcommand{\beqa}{\begin{eqnarray}}
\newcommand{\eeqa}{\end{eqnarray}}
\newcommand{\eps}{\epsilon}
\newcommand{\om}{\omega}

\def\bea{\begin{eqnarray}}
\def\eea{\end{eqnarray}}
\def\be{\begin{equation}}
\def\ee{\end{equation}}

\begin{document}

\title{Finite size effects for the gap in the excitation spectrum of the one-dimensional Hubbard model}
\author{M. Colom\'e-Tatch\'e $^{1,2}$, S. I. Matveenko$^{1,3}$ and G. V. Shlyapnikov$^{1,4}$}
\affiliation{\mbox{$^1$ Laboratoire de Physique Th\'eorique et Mod\'eles Statistiques, Universit\'e Paris Sud, CNRS,}\\\mbox{91405~Orsay, France.} \\
\mbox{$^2$Institute for Theoretical Physics, Leibniz Universit\"at Hannover, Appelstr. 2, D-30167, Hannover, Germany.}
\mbox{$^3$Landau Institute for Theoretical Physics, Kosygina Str. 2, 119334, Moscow, Russia.}
\mbox{$^4$ Van der Waals-Zeeman Institute, University of Amsterdam, Valckenierstraat 65/67,}\\\mbox{1018 XE Amsterdam, The Netherlands.}
}
\date{\today}

\begin{abstract}
We study finite size effects for the gap of the  quasiparticle excitation spectrum in the weakly interacting regime
one-dimensional Hubbard model  with on-site attraction. Two type of corrections to the result of the
thermodynamic limit are obtained. Aside from a power law (conformal) correction due to
gapless excitations which behaves as $1/N_a$, where $N_a$ is the number of
lattice sites, we obtain corrections related to the existence of gapped
excitations. First of all, there is  an exponential correction which in the
weakly interacting regime ($|U|\ll t$) behaves as $\sim \exp (-N_a \Delta_{\infty}/4 t)$ in
the extreme limit of $N_a \Delta_{\infty} /t \gg 1$,
where $t$ is the hopping amplitude, $U$ is the on-site energy, and  $\Delta_{\infty}$ is the  gap in the
thermodynamic limit. Second, in a finite size system a spin-flip producing
unpaired fermions leads to the appearance of solitons with non-zero momenta, which provides an extra (non-exponential) contribution $\delta$.
For moderate but still large values of $N_a \Delta_{\infty} /t$, these corrections significantly increase and may become comparable with the 
$1/N_a$ conformal correction. Moreover, in the case of weak interactions where $\Delta_{\infty}\ll t$, the exponential correction exceeds higher 
order power law corrections in a wide range of parameters, namely for $N_a\lesssim (8t/\Delta_{\infty})\ln(4t/|U|)$, and so does $\delta$ even in 
a wider range of $N_a$.
For sufficiently small number of particles,  which can be  of the order of thousands in the weakly
interacting  regime, the gap is fully dominated by finite size effects.
\end{abstract}


\maketitle

\section{Introduction}
 
The one-dimensional (1D)  Hubbard model with  only on-site interaction is exactly solvable by the Bethe
 Ansatz, and the properties of this model have been widely studied \cite{book}.
 Low lying excitations of the Hubbard model with attraction are gapless charge excitations and gapped
 spin excitations, whereas for the  repulsive model  the gap exists only at half filling in the charge sector.

Finite size corrections to the ground state energy $E_0$, due to the gapless part of the spectrum, follow
from  conformal field theory  \cite{BPZ,Cardy,Blote-Cardy-Nightingale,Affleck} and have the form:
 $E_0 - \epsilon _0 L=-\pi v \hbar /6L$,  where $L = N_a  a$ is the size of the system,
$N_a$ is the number of lattice sites, $a$  is the  lattice constant, $\epsilon_{0}$ is the energy  per
unit length  in the thermodynamic limit, and $v$ is the velocity of gapless excitations.
Finite size corrections to the energies of low-lying gapless excitations are also proportional to
$v/N_a$, and
 the proportionality coefficient depends on the scaling dimensions of the primary fields. However, finite
 size corrections originating from the  gapped sector remained unknown for the Hubbard model.
Finite size effects are expected to be
 important for  sufficiently small systems, such as cold atoms in a 1D optical lattice, where the
 number of particles and lattice sites ranges from several tens to several hundreds \cite{xxx,xxx1}.

In this paper we solve the Bethe Ansatz equations for a finite number of particles and
calculate  finite size corrections to the  gap for the attractive  Hubbard model.
As expected, there are  power law $1/N_a $ corrections due to gapless excitations, and
we also find contributions related to the existence of gapped excitations.
First of all, there is  an exponential correction which in the
weakly interacting regime ($|U|\ll t$) behaves as $\sim \exp (-N_a \Delta_{\infty}/4 t)$ in
the extreme  limit of $N_a \Delta_{\infty} /t \gg 1$,
where $t$ is  the  hopping amplitude, $U$ is the on-site interaction, and $\Delta_{\infty}$ is the  gap in the
thermodynamic limit. Second, in a finite size system a spin-flip producing
unpaired fermions leads to the appearance of solitons with non-zero momenta,
which provides an extra (non-exponential) contribution. For moderate but still large values of $N_a \Delta_{\infty} /t$, these corrections may become comparable with the 
$1/N_a$ conformal correction. Moreover, in the case of weak interactions where $\Delta_{\infty}\ll t$, the exponential correction exceeds higher 
order power law corrections in a wide range of parameters, namely for $N_a\lesssim (8t/\Delta_{\infty})\ln(4t/|U|)$, and so does $\delta$ even in 
a wider range of $N_a$.
We  find that the value of the gap increases with decreasing the system size and show how the gap becomes dominated by finite size effects in sufficiently small systems.

The paper is organized as follows. In section II we introduce the Hubbard model
together with related Bethe Ansatz equations,  and discuss the thermodynamic limit.
In section III we present a general  approach for finding
finite size corrections to the ground state energy  and to the  gap
and discuss the structure  of the gap.
Section IV  contains our  results for corrections due to the gapped sector at half
filling, and section V the results for power law corrections.
In Section VI we discuss our numerical and analytical results, and in Section
VII present for completeness a perturbative approach for solving the Bethe Ansatz equations in the limiting case of $L \ll
a t /U$ . In Section VIII we conclude.

\section{General equations, thermodynamic limit}

The Hubbard model for  a system of  interacting  spin-1/2 fermions  on a lattice is described  by the
 Hamiltonian
\be
\label{Hamiltonian_Hubbard_Model}
H=-t\sum_{\sigma=\uparrow,\downarrow;j=1}^{N_a}(c_{j, \sigma}^{\dagger}c_{j + 1, \sigma}+c_{j + 1, \sigma}^{\dagger}c_{j, \sigma})+U\sum_{j=1}^{N_a}n_{j,\uparrow}n_{j,\downarrow},
\ee
where  the subscript $j$ labels the lattice sites. The index $\sigma$ labels the spin  projection,
 $c^{\dagger}_{j, \sigma}$ and $c_{j, \sigma}$ are the creation and annihilation fermion operators,
 and $n_{j,\sigma}=c^{\dagger}_{j,\sigma}c_{j, \sigma}$  are the particle number operators.
Below we express all quantities having the dimension  of energy in units of $t$, and quantities having the dimension of length in units of the lattice constant $a$.

Lieb and Wu  have solved the Fermi-Hubbard model by means of the Bethe Ansatz \cite{lieb-wu}. The
corresponding eigenvalue equations read
\beqa
\label{Discrete_Bethe_Ansatz_Hubbard_Model}
\sum_{j = 1}^N 2 \arctan \left( \frac{\lambda_{\alpha} - \sin k_j }{u}\right) = 2 \pi J_{\alpha} +
\sum_{\beta = 1}^M 2 \arctan \left( \frac{\lambda_{\alpha} - \lambda_{\beta}}{2u}\right) ,
\quad \alpha = 1, ...M, \\
N_a k_j = 2 \pi I_j - \sum_{\beta = 1}^M 2 \arctan \left( \frac{\sin k_j - \lambda_\beta }{u}\right),
\quad j = 1, ...N ,
\label{dba2}
\eeqa
where $u=|U|/4t$, $M$  is  the number  of spin-down fermions,  and $N$ is the  total number of particles.
The energy of a given state is expressed through  the charge momenta $k_j$:
\beq
\label{energy-discrete}
E_N = -2 \sum_{j= 1}^N \cos k_j ,
\eeq
and   it  depends on the spin rapidities $\lambda_{\alpha}$ only implicitly through the coupled equations (\ref{Discrete_Bethe_Ansatz_Hubbard_Model}) and (\ref{dba2}).
For the ground state the quantum numbers $I_j$ and $J_{\alpha}$ are integers or half-odd integers depending on the parities of $N$ and $M$:
\bea
\label{Quantum_Numbers_Hubbard_Model_Ground_State}
J_{\alpha} =  \frac{N+M+1}{2}(\textrm{mod } 1), \qquad I_{j} =\frac{M}{2}(\textrm{mod }1) .
\eea

For the Hubbard model with attraction ($U < 0$), there is a gap in the spectrum of spin excitations.
Considering a finite number of particles we define the gap  $\Delta$  as:
\begin{equation}
2 \Delta = E_{N +2 }(N_{\uparrow} + 2, N_{\downarrow}, U) + E_{N -2}(N_{\uparrow}-2, N_{\downarrow}, U) - 2 E_{N}(N_{\uparrow}, N_{\downarrow}, U),
\label{2delta}
\end{equation}
where $E_{N}(N_{\uparrow}, N_{\downarrow};U)$ is the ground state energy for a system with
 $N_{\uparrow}$ spin-up  and $N_{\downarrow} = N - N_{\uparrow}$ spin-down fermions in a lattice with $N_a$ lattice sites,
 at the  interaction strength $U$.  This definition is convenient as it does not change the parity of
 the quantum numbers $I_j$ and $J_{\alpha}$.
Without loss of generality, we may put $N_{\downarrow}\leq N_{\uparrow}$.

In the thermodynamic limit, where $N \to \infty$ and $N_a \to \infty$ while keeping constant densities
$n = N/L$ and $n_{\downarrow } =N{\downarrow } /L$,  Eq.  (\ref{2delta}) leads to the same result as the definition
$ \Delta=E_{N_a}(N_{\uparrow}+1,N_{\downarrow};U)- 2
E_{N_a}(N_{\uparrow},N_{\downarrow};U) + E_{N_a}(N_{\uparrow}-1,N_{\downarrow};U)$
introduced  in Refs. \cite{kr-ovch,lieb-wu}.

In the thermodynamic limit the density of momenta $k$ and the density of rapidities $\lambda$
are defined as $\rho(k)=L^{-1}{\partial I}/{\partial k}$ and
$\sigma(\lambda)=L^{-1}{\partial J}/{\partial\lambda}$, respectively. Then, the Bethe Ansatz equations
for the ground state  of the repulsive model  become:
\begin{eqnarray}
\label{Thermodynamic_Bethe_Ansatz_Hubbard_Model-k}
\rho_{\infty}(k)&=&\frac{1}{2\pi}+\frac{\cos k}{\pi} \int_{-B}^{B} \frac{u}{u^2 +
(\lambda-\sin k)^2} \sigma_{\infty}(\lambda) d\lambda,\\
\label{Thermodynamic_Bethe_Ansatz_Hubbard_Model-lambda}
\sigma_{\infty}(\lambda)&+&\frac{1}{\pi}\int_{-B}^B \frac{2u }{4u^2+(\lambda-\lambda')^2}
\sigma_{\infty}(\lambda') d\lambda'=\frac{1}{\pi}\int_{-Q}^Q \frac{u}{u^2+(\lambda-\sin k)^2}\rho_{\infty}(k) dk,
\end{eqnarray}
where the constants $B$ and $Q$ are given by
\be
\label{Number_Particles_Hubbard_Model}
\frac{N}{N_a} =n=\int_{-Q}^Q \rho_{\infty}(k)dk,\qquad
\frac{N_{\downarrow}}{N_a} =n_{\downarrow}=\int_{-B}^B \sigma_{\infty}(\lambda) d\lambda,
\ee
and the ground state energy follows from the relation:
\beq
E_{\infty} = - 2 N_a \int_{-Q}^Q \rho_{\infty} (k) \cos k \, \, dk.
\eeq
The value of the spin gap  for  an arbitrary filling factor
 has been calculated in the thermodynamic limit in Refs. \cite{kr-ovch,lar}. In the case of weak attraction the result is
\be
 \Delta_{\infty} =\frac{16 \sin^{3/2}(\pi n /2) \sqrt{u}}{\pi}\exp
  \left(- \frac{\pi\sin\left({\pi n}/{2}\right)}{2u}\right),
\label{ddd}
\ee
and the validity of Eq.  (\ref{ddd}) requires large values of the exponent.

The particle-hole symmetry and the symmetry with respect to interchanging spin-up and spin-down  fermions
allow one to establish relations between ground state energies of the repulsive and attractive Hubbard models \cite{lieb-wu}:
\beqa
E_{}(N_{\uparrow},N_{\downarrow};U)= -(N_a-N_{\uparrow}-N_{\downarrow})U+E_{}(N_a-N_{\uparrow},N_a-N_{\downarrow};U)  \nonumber \\
=N_{\uparrow}U+E_{}(N_{\uparrow},N_a-N_{\downarrow};-U) =  N_{\downarrow}U+E_{}(N_a-N_{\uparrow},N_{\downarrow};-U).
\label{Symmetries_Hubbard_Model}
\eeqa

Using  Eqs.~(\ref{2delta}) and (\ref{Symmetries_Hubbard_Model}) we can express the spin gap for the attractive Hubbard model through the energies of the repulsive model:
\be
2\Delta = 2|U| + E_{N_a -2}(M - 2,N_a -M, |U|)+ E_{N_a - 2}(N_a - M -2, M, |U|)-2 E_{N_a} (M, N_a -M, |U|) .
\label{Gap_Hubbard_Model_2}
\ee
For the half-filled case  ($N =2 M = N_a$),  with  $N_{\uparrow}=N_{\downarrow}=M$,
Eq.  (\ref{Gap_Hubbard_Model_2}) takes the form:
\be
\Delta = |U| + E_{N_a -2}(N_a/2,N_a/2 -2, |U|) -E_{N_a} (N_a/2, N_a/2, |U|).
\label{Gap_Hubbard_Model3}
\ee

Below we calculate the gap for the attractive Hubbard model. For this purpose we first perform  calculations of ground state energies for the repulsive Hubbard
model, where the momenta $k_j$ are real numbers, and then obtain $\Delta$ for the attractive model by using Eq. (\ref{Gap_Hubbard_Model_2}) (Eq. (14) for
the half filled case).  This allows us to find exponential  finite size corrections, which is  not possible  in direct calculations for the case of attraction  where $k_j$ are
complex and can be found only with an exponential accuracy.

\section{Finite size corrections. General approach}
Thus, in order to calculate  finite size corrections to the gap we have  to obtain
the three energies of a finite size system,  entering the right hand side of Eq. (\ref{Gap_Hubbard_Model_2}).
We will follow the scheme proposed by   de Vega and  Woynarovich \cite{w2}, which introduces a formalism  allowing us to use the Bethe Ansatz in order to calculate  finite size
corrections to the energy of the ground state. The scheme consists of writing  the Bethe Ansatz equations (\ref{Discrete_Bethe_Ansatz_Hubbard_Model}) and (\ref{dba2}) in the form:
\bea
\label{zs}
Z^s (\lambda) &=& \frac{1}{N_a}\sum_{j}^N \frac{1}{\pi}\arctan\frac{\lambda -\sin k_j}{u}-\frac{1}{N_a}\sum_{\beta}^M \frac{1}{\pi}\arctan\frac{\lambda -\lambda_{\beta}}{2u}, \quad
Z^s (\lambda_{\alpha}) = \frac{J_{\alpha}}{N_a}, \\
Z^c(k) &=& \frac{k}{2\pi} + \frac{1}{N_a}\sum_{\alpha}^M \frac{1}{\pi}\arctan\frac{\sin k -\lambda_{\alpha}}{u}, \quad Z^c (k_j) = \frac{I_j}{N_a}.
\label{zc}
\eea
We then define the densities of momenta $k$ and rapidities $\lambda$ for a finite size system as
\bea
\sigma_N (\lambda)  &\equiv& \frac{dZ^s}{d\lambda}= \frac{1}{2\pi N_a} \sum_{j = 1}^N
K_1 ( \lambda - \sin k_j) -\frac{1}{2\pi N_a} \sum_{\beta = 1}^M
K_2 (\lambda - \lambda_{\beta}), \label{rna}\\
\rho_N (k) &\equiv& \frac{dZ^c}{dk} = \frac{1}{2\pi} + \frac{1}{N_a}\frac{\cos k}{2\pi} \sum_{\alpha = 1}^M
K_1 (\sin k- \lambda_{\alpha}),
\label{sna}
\eea
where $K_1 (x) = {2u}/({u^2 + x^2})$, and $ K_2 = {4u}/({4u^2 + x^2})$. The densities
 satisfy the relations:
\be
\int_{\Lambda_-}^{\Lambda_+} \sigma_N (\lambda )d \lambda = \frac{M}{N_a}, \quad
\int_{Q_-}^{Q_+} \rho_N (k) d k = \frac{N}{N_a},
\ee
where $Q_{\pm }$, $\Lambda_{\pm }$ are determined from the equations
\be
\label{Q_Lambda}
Z^c (Q_+) = \frac{I_+}{N_a} = \frac{I_{max} + 1/2}{N_a}; \;\; Z^s (\Lambda_+) = \frac{J_+}{N_a} = \frac{J_{max} + 1/2}{N_a}.
\ee

We first perform calculations for the half-filled case. According to Eq. (\ref{Gap_Hubbard_Model3}) we have to calculate $E_{N_a} (N_a /2, N_a /2;|U|)$ and  $E_{N_a} (N_a /2 ,
N_a /2 - 2;|U|)$. In the  former case we have $N = N_a$ and $N_{\uparrow} = N_{\downarrow} = N_a /2 = M$, and the quantum numbers for the ground state are
\bea
\label{quantum-numbers}
J_{\alpha} &=& \{ -\frac{M -1}{2} ,..., -1, 0, 1, ..., \frac{M -1}{2} \},\nonumber\\
I_j &=& \{ -\frac{N -1}{2} ,..., -\frac{1}{2}, \frac{1}{2}, ..., \frac{N -1}{2} \}.
\eea
With $N=N_a$ and $M=N_a/2$, from Eq. (\ref{quantum-numbers}) we have
\be
J_{max} = \frac{N_a /2 - 1}{2} ;
\quad  I_{max} = \frac{N_a - 1}{2}.
\ee
Then, from Eq. (\ref{Q_Lambda}) we obtain
\bea
 J_+ = \frac{N_a}{4}, \quad Z_s (\Lambda_+ ) =  \frac{1}{4},\nonumber\\
 I_+ = \frac{N_a}{2}, \quad Z_c (Q_+ ) = \frac{1}{2},
\label{IJ}
\eea
and Eqs. (\ref{zs}) and (\ref{zc}) lead to
\be
\Lambda_+ = \infty; \qquad Q_+ = \pi.
\ee

When calculating the energy $E_{N_a} (N_a /2, N_a /2 - 2; |U|)$ we have $N = N_a -2$,
$N_{\uparrow} = N_a /2 $, and $N_{\downarrow} = N_a /2 - 2$.
It is convenient to introduce two additional  particles
with momenta $k_h=\{k_1, \, k_{N_a}\}$ and spin rapidities
 $\lambda_h =
\{\lambda_1, \, \lambda_{N_a /2}\}$ in order to
 satisfy the conditions $Q_{\pm} = \pm \pi$ and $\Lambda_{\pm} = \pm \infty$.
The discrete Bethe Ansatz equations in this case read
\bea
\sum_{j = 1}^{N_a} 2 \arctan \left(\frac{\lambda_{\alpha} - \sin k_j }{u} \right) =
2 \pi J_{\alpha} +\sum_{\beta = 1}^{N_a /2} 2 \arctan  \left( \frac{\lambda_{\alpha} - \lambda_{\beta}}{2u}\right)\nonumber\\
+\sum_{k_h} 2 \arctan \left(\frac{\lambda_{\alpha} - \sin k_h}{u}\right) - \sum_{\lambda_h} 2\arctan \left(\frac{\lambda_{\alpha} - \lambda_h}{2u}\right), \quad \alpha =
1, ...N_a/2, \\
N_a k_j = 2 \pi I_j - \sum_{\beta = 1}^{N_a/2} 2 \arctan \left(\frac{\sin k_j - \lambda_\beta }{u}\right)+ \sum_{\lambda_h} 2 \arctan \left(\frac{\sin k_j - \lambda_h }{u}\right),
\quad j = 1, ...N_a .
\eea
Note that we added  only  4 additional equations   for defining the numbers $k_h$, $\lambda_h$.
Other equations are exactly the Bethe Ansatz equations for  $N_a - 2 $ particles. The sets  ${J_{\alpha}}$, ${I_j}$ are the same as for the ground state of $N_a$ particles
(\ref{quantum-numbers}).

For the case $N = N_a$  we  rewrite  equations (\ref{rna}) and(\ref{sna}) in the form of
 integral equations for the densities $\sigma_{N_a} (\lambda )$, $\rho_{N_a} (k)$:
\bea
\label{Densities_Finite_N_Hubbard_Model}
\sigma_{N_a}(\lambda) &=& \frac{1}{2\pi} \int_{-\pi}^{\pi} K_1(\lambda - \sin k) \rho_{N_a}(k)dk - \frac{1}{2\pi} \int_{-\infty}^{\infty} K_2(\lambda - \mu) \sigma_{N_a}(\mu)d\mu
\nonumber
\\
&+&\frac{1}{2\pi}\int_{-\pi}^{\pi} K_1(\lambda -\sin k) X^c_{N_a}(k) dk - \frac{1}{2\pi}\int_{-\infty}^{\infty} K_2(\lambda -\mu) X_{N_a}^s(\mu)d\mu, \label{dfl} \\
\rho _{N_a}(k) &=& \frac{1}{2\pi} + \frac{1}{2\pi} \int_{-\infty}^{\infty} K_1(\sin k -\lambda) \cos k\; \; \sigma_{N_a}(\lambda )d\lambda\nonumber\\
&+& \frac{1}{2\pi} \int_{-\infty}^{\infty} \cos k \; \; K_1(\sin k - \lambda ) X_{N_a}^s(\lambda)d\lambda,
\label{dfr}
\eea
with the quantities $X_{N_a}^s$ and $X_{N_a}^c$ defined as
\bea
\label{20}
X_{N_a}^c (k) &=& \frac{1}{N_a}\left[ \sum_{j = 1}^{N_a} \delta(k - k_j) \right] - \rho_{N_a} (k), \\
X_{N_a}^s (\lambda) &=& \frac{1}{N_a}\left[ \sum_{\alpha = 1}^{N_a /2} \delta(\lambda - \lambda_{\alpha})
\right] - \sigma_{N_a} (\lambda).
\label{21}
\eea
For $N = N_a -2$ we do the same and  obtain
\bea
\sigma_{N_a -2} (\lambda) &=& \frac{1}{2 \pi} \int_{-\pi}^{\pi} K_1 ( \lambda - \sin k ) \rho_{N_a -2} (k ) \; dk - \frac{1}{2 \pi} \int_{-\infty}^{\infty} K_2 ( \lambda - \mu )
\sigma_{N_a -2} (\mu ) \;
d \mu  \nonumber\\
&+&\frac{1}{2 \pi}\int_{-\pi}^{\pi}   K_1 ( \lambda -\sin k ) X^c_{N_a -2} (k ) \; dk - \frac{1}{2 \pi}\int_{-\infty}^{\infty}  K_2 ( \lambda -\mu ) X_{N_a -2}^s (\mu ) \; d \mu
\nonumber\\
&-& \frac{1}{2 \pi N_a}\sum_{k_h} K_1 ( \lambda - \sin k_h  ) + \frac{1}{2 \pi N_a}\sum_{\lambda_h} K_2 (\lambda - \lambda_h ),  \label{sN} \\
\rho _{N_a -2} (k) &=& \frac{1}{2 \pi} + \frac{1}{2 \pi} \int_{-\infty}^{\infty} K_1 (\sin k - \lambda )  \sigma_{N_a -2} (\lambda ) \cos k \;  d \lambda  \nonumber\\
 &+& \frac{1}{2 \pi}\int_{-\infty}^{\infty}   K_1 (\sin k - \lambda ) X_{N_a -2}^s (\lambda )\; \cos k \; d \lambda
  -\frac{1}{2 \pi N_a}\sum_{\lambda_h} K_1 (\sin k - \lambda_h ) \cos k .
\label{rN}
\eea
The functions $X_{N_a -2}^{s,c}$ are given by
\beq
X^s_{N_a -2} (\lambda) = \frac{1}{N_a} \sum_{\alpha =1}^{N_a /2} \delta(\lambda - \lambda_{\alpha}) -\sigma_{N_a -2}(\lambda),
\label{x1}
\eeq
\beq
 X^c_{N_a -2} (k) = \frac{1}{N_a} \sum_{j=1}^{N_a} \delta(k - k_j) -\rho_{N_a -2}(k),
\label{x2}
\eeq
 where   the summation over   $k_j$, $\lambda_{\alpha}$ includes the additional numbers
$k_h$, $\lambda_h$.

We  first consider the thermodynamic limit where the terms $X^{s, c}$ in Eqs. (\ref{Densities_Finite_N_Hubbard_Model}), (\ref{dfr}), (\ref{sN}) and
(\ref{rN}) vanish. For the half-filled case  of $N = N_a$ the solution is known \cite{lieb-wu}:
\bea
\label{Densities_Hubbard_Model_Infinity}
\rho_{\infty, N_a}(k) &=& \frac{1}{2\pi} + \frac{\cos k}{2\pi} \int_0^{\infty}\frac{\cos(\om \sin k)
\exp (-\om u)}{\cosh \om u}J_0(\om) d\om , \\
\sigma_{\infty, N_a} (\lambda) &=& \frac{1}{2\pi} \int_0^{\infty} \frac{\cos \om \lambda}{\cosh \om u} J_0(\om) d \om.
\label{dhbm}
\eea
Integrating the density of momenta over $dk$  we obtain
\be
Z_{\infty, N_a}^c (k) = \frac{k}{2\pi} + \frac{1}{2\pi}\int_0^{\infty} \frac{\sin (\om \sin k )
\exp (-\om u)}{\om \cosh \om u} J_0(\om) d \om.
\label{zn}
\ee

For the case  of $N = N_a -2$ the solution is easily found:
\bea
\label{Densities_Hubbard_Model_Na-2_Infinity-lambda}
\sigma_{\infty, N_a -2} &=& \sigma_{\infty, N_a} - \frac{1}{2 \pi N_a} \sum_{k_h} \frac{\pi}{2 u}\frac{1}{\cosh [{\pi}(\lambda - \sin k_h )/2u]}\nonumber\\
&+& \frac{1}{ 2\pi N_a} \sum_{\lambda_h} \int_0^{\infty}\frac{\cos [\om (\lambda -\lambda_h )]}{\cosh \om u} \exp (- \om u ) \; d \om, \\
\label{Densities_Hubbard_Model_Na-2_Infinity-k}
\rho_{\infty, N_a -2} &=& \rho_{\infty, N_a} - \frac{\cos k}{ 2\pi N_a}\sum_{k_h }  \int_0^{\infty}\frac{\cos [\om (\sin k -\sin k_h )]}{\cosh \om u} \exp (- \om u) \; d \om
\nonumber\\
&-& \frac{\cos k}{2 \pi N_a} \sum_{\lambda_h } \frac{\pi}{2 u} \frac{1}{\cosh [{\pi}(\sin k -  \lambda_h )/2u]} ,
\label{dhmn-2}
\eea
and
\bea
Z_{\infty, N_a -2}^c (k ) &=& Z_{\infty, N_a}^c (k) - \frac{1}{2\pi N_a}\sum_{k_h}\int_0^{\infty}
\frac{\sin \left[\om (\sin k -\sin k_h)\right]  \exp (-\om u)}{\om \cosh \om u}
 d \om \nonumber\\
&-& \frac{1}{2 \pi N_a} \sum_{\lambda_h} 2 \arctan \left[\tanh \left(\frac{\pi}{4 u}(\sin k - \lambda_h )\right)\right].
\label{zn-2}
\eea

For calculating the finite size corrections we should find the differences between the densities $\rho _{N_a} (k)$,
$\sigma_{N_a} (\lambda)$, $  \sigma_{N_a -2} (\lambda)$, $\rho _{N_a -2} (k)$ and their thermodynamic limit values.
 Subtracting  equations of the thermodynamic limit  from Eqs. (\ref{dfl}), (\ref{dfr})
   for the case $N = N_a$ we obtain:
\begin{eqnarray}
\label{31}
\delta \sigma_{N_a} (\lambda)&=&\sigma_{N_a} (\lambda)-\sigma_{\infty,{N_a}}(\lambda ) = \frac{1}{2 \pi} \int_{-\pi}^{\pi} K_1 ( \lambda - \sin k ) \delta \rho_{N_a} (k ) \;
dk -\frac{1}{2 \pi}
\int_{-\infty}^{\infty} K_2 ( \lambda - \mu )  \delta \sigma_{N_a} (\mu ) \; d \mu   \nonumber \\
& &+\frac{1}{2 \pi}\int_{-\pi}^{\pi}  K_1 ( \lambda -\sin k ) X^c_{N_a} (k )\; d k - \frac{1}{2 \pi}\int_{-\infty}^{\infty} K_2 ( \lambda -\mu ) X_{N_a}^s (\mu ) \; d \mu , \\
\delta\rho _{N_a} (k) &=&\rho_{N_a} (k)-\rho_{\infty,{N_a}}(k) =  \frac{1}{2 \pi} \int_{-\infty}^{\infty}
K_1 (\sin k - \lambda ) \cos k  \; \delta \sigma_{N_a} (\lambda ) \; d \lambda \nonumber \\
 & &+ \frac{1}{2 \pi}\int_{-\infty}^{\infty}  \cos k  \; K_1 (\sin k - \lambda ) X_{N_a}^s (\lambda ) \;
 d \lambda.
\label{32}
\end{eqnarray}
 Equations  (\ref{31}) and(\ref{32})  lead to algebraic equations for the Fourier transforms of $ \delta \sigma_{N_a} (\lambda)$ and $\delta\rho _{N_a} (k) $, which yields:
\bea
\delta \sigma_{N_a} (\lambda ) &=& \int_{-\pi}^{\pi}  \frac{\pi}{2u}\frac{1}{\cosh \left[ {\pi}(\lambda-\sin k)/2u \right]} X^c_{N_a} (k) \frac{d k}{2\pi}\nonumber\\
&-&\int_{-\infty}^{\infty} d \mu \int_0^{\infty} \frac{\cos [\om (\lambda - \mu)]}{\cosh (\om u)} \exp (- \om u) X_{N_a}^s (\mu )  \frac{d \om}{2 \pi}.
\label{drds} \\
\label{Delta_Rho_Delta_Sigma}
\delta\rho _{N_a} (k) &=& \cos k \int_0^{\infty} \frac{d \om}{2 \pi} \int_{-\pi}^{\pi}  \frac{\exp (- \om u)}{\cosh (\om u)} \cos [\om (\sin k - \sin q )] X^c_{N_a} (q) \; d q
\nonumber
\\
&+& \frac{\cos k}{2 \pi} \int_{- \infty}^{\infty}   \frac{\pi}{2 u}
\frac{1}{\cosh  \left[{\pi} (\sin k - \lambda )/2u \right] } X_{N_a}^s (\lambda ) \; d \lambda .
\eea
Equations for $\delta \sigma_{N_a-2} (\lambda) $ and $\delta\rho _{N_a -2} (k)$
are  obtained in a similar way  taking into account that in the thermodynamic limit $k_h = \pm \pi$ and
 $\lambda_h = \pm \infty$. This gives Eqs. (\ref{31}), (\ref{32}) and (\ref{drds}),
 (\ref{Delta_Rho_Delta_Sigma}) where $N_a $ is replaced by $N_a -2$ and $X^{s, c}$ by $\tilde{X}^{s, c}$.
 The quantity $\tilde{X}^{s}_{N_a -2}$
is given by  Eq.(\ref{x1}) in which the values of $\lambda_h$ are put equal to $+\infty$ and $-\infty$:
\be
 \tilde{X}^s_{N_a -2} (\lambda) = \frac{1}{N_a} \sum_{\alpha =1}^{N_a /2 -2} \delta(\lambda - \lambda_{\alpha})
 -\sigma_{N_a -2}(\lambda),
\ee
and $\tilde{X}^{c}_{N_a -2}$ by Eq. (\ref{x2}) where the values $k_h$ are put equal to $+\pi$ and $-\pi$:
\be
 \tilde{X}^c_{N_a -2} (k) = \frac{1}{N_a}\left[ \sum_{j=1}^{N_a -2} \delta(k - k_j)
+\delta (k -\pi) +\delta (k + \pi )\right]  -\rho_{N_a -2}(k).
\ee

The ground state energies   (\ref{energy-discrete}) for the  considered states  can be  rewritten as
\bea
\label{en}
\frac{E_{N_a}}{N_a} &=& -2\int_{-\pi}^{\pi} \rho_{N_a} (k)   \cos k \; d k - 2 \int_{-\pi}^{\pi}
 X^c_{N_a} (k) \cos k \; d k , \\
\frac{E_{N_a -2}}{N_a} &=& -2\int_{-\pi}^{\pi} \rho_{N_a - 2} (k) \cos k \; d k - 2 \int_{-\pi}^{\pi}
 {X}^c_{N_a - 2} (k) \cos k \; d k   +\frac{ 2}{N_a}\sum_{k_h} \cos k_h \nonumber \\
 &=& -2\int_{-\pi}^{\pi} \rho_{N_a - 2} (k) \cos k \; d k - 2 \int_{-\pi}^{\pi}
 \tilde{X}^c_{N_a - 2} (k) \cos k \; d k  - \frac{ 4}{N_a}.
\label{en-2}
\eea

The finite size corrections to the  ground state energy are given by
\be
\label{Delta_Energy}
\delta E_N=E_N-E_{\infty,N},
\ee
where  the energies $E_{\infty,N}$ in the thermodynamic limit are
\beq
\frac{E_{\infty, N_a}}{N_a} = -2  \int_{-\pi}^{\pi} \rho_{\infty, N_a}(k) \cos k \;  dk,
\eeq
\beq
 \frac{E_{\infty, N_a -2}}{N_a} = -2 \int_{\pi}^{\pi} \rho_{\infty, N_a -2}(k) \cos k \;  dk  -\frac{4}{N_a},
\label{EN}
\eeq
with $\rho_{\infty, N_a}(k)$  given by Eq. (\ref{Densities_Hubbard_Model_Infinity}) and    $\rho_{\infty, N_a -2}(k)$
by Eq. (\ref{dhmn-2})  with  $k_h = \pm \pi$ and $\lambda_h = \pm \infty$.

Using Eqs.   (\ref{en}), (\ref{en-2}), (46), (\ref{Delta_Rho_Delta_Sigma}), and Eq. (\ref{EN}) with
$\rho_{\infty , N_a}$ and $\rho_{\infty , N_a -2}$ following from Eqs. (\ref{Densities_Hubbard_Model_Infinity}) and Eq. (\ref{dhmn-2}),
we obtain
\be
\frac{\delta E_{N_a}}{N_a} = -\int_{-\pi}^{\pi} \eps_c (k) X^c_{N_a} (k) \;dk - \int_{- \infty}^{\infty}\eps_s (\lambda ) X^s_{N_a} (\lambda )\;d \lambda
\label{deltae},\
\ee
\be
\frac{\delta E_{N_a -2}}{N_a} = -\int_{-\pi}^{\pi} \eps_c (k) \tilde{X}^c_{N_a -2} (k) \;dk - \int_{- \infty}^{\infty}\eps_s (\lambda ) {X}^s_{N_a -2} (\lambda )\;d \lambda,
\label{deltae2}\
\ee
where
\bea
\label{epsc}
\eps_c (k) &=& 2 \cos k  +2\int_0^{\infty} \frac{J_1(\om)\exp (- \om u )}{\om \cosh (\om u) }\cos(\om \sin k) \;d \om , \\
\eps_s (\lambda) &=& 2  \int_0^{\infty} \frac{J_1(\om) }{ \om \cosh (\om u) }\cos(\om \lambda) \;d \om,
\label{epss}
\eea
and we used the fact that $\eps_s (\pm \infty ) = 0$, so that $\tilde{X}_{N_a -2}^s (\lambda )$ can
be replaced by  $X_{N_a -2}^s (\lambda )$ in Eq. (\ref{deltae2}).

For the repulsive Hubbard model at half filling, the gap is in the charge sector and the spin sector is gapless. Accordingly,
the first term in the right-hand side of Eq. (\ref{deltae}) describes the contribution of gapped
charge excitations, and the second term is due to the contribution of gapless spin excitations.

We now return to Eq. (\ref{Gap_Hubbard_Model3}) and using Eqs. (\ref{deltae}), (\ref{deltae2})
 write the finite size corrections to the gap of the attractive case in the form
\be
\delta \Delta = \delta E_{N_a -2} - \delta E_{N_a }= \delta \Delta_{ng} + \delta \Delta_{g} + \delta,
\label{g00}
\ee
where
\beq
\delta \Delta_{ng} /N_a = \int_{-\infty}^{\infty} \eps_s (\lambda) X^s_{N_a}\;d \lambda- \int_{-\infty}^{\infty}
\eps_s (\lambda) {X}^s_{N_a -2 }\;\, d \lambda,
\label{gss}
\eeq
is the contribution of the gapless  sector,
\beq
\delta \Delta_g /N_a =\int_{-\pi}^{\pi} \eps_c (k) X^c_{N_a}\;dk - \int_{-\pi}^{\pi}\eps_c (k) {X}^c_{N_a -2 }\;\, dk ,
\label{Gap_Spin_Sector}
\eeq
is due to gapped excitations, and
we used  a relation
\[
 \tilde{X}^c_{N_a -2} =
 X_{N_a -2}^c - \frac{1}{N_a} \sum_{h =1}^2 \delta (k - k_h) + \frac{1}{N_a} \left[\delta (k - \pi) + \delta (k + \pi)\right].
\]
The term $\delta$ which is  present in the gapped sector is given by
\beq
 {\delta} = \sum_h \eps (k_h) - \eps (\pi) - \eps (-\pi)
\label{g0}
\eeq
and for $N_a\Delta_{\infty}\gtrsim 1$ it reduces to
$$\delta\approx \eps_c^{\prime\prime} (\pi) (k_h^+ - \pi )^2,$$
where $k_h^+$ is the value of $k_h$ which is close to $\pi$ at large $N_a$.
This term behaves as $1/N_a^2$ for large $N_a$.  In the limit of $u \ll 1$ the value of $k_h^+$ can be found from Eqs. (\ref{zn}) and (\ref{zn-2}) using the condition $Z_{N_a -2}
=1/2 - 1/2N_a$.
For $N_a\Delta_{\infty}\gg 1$ we then obtain
\begin{equation}              \label{deltass}
\frac{\delta}{\Delta_{\infty}} = 2\left(\frac{2 \pi}{N_a\Delta_{\infty}}\right)^2,
\end{equation}
where at half filling the gap of the thermodynamic limit $\Delta_{\infty}$ is given by 
\beq        
\Delta_{\infty} = 4 u  -4 + 4\int_0^{\infty} \frac{\exp (-\om u) J_1 (\om )}{\om \cosh \om u}d \om
\label{h0}
\eeq
at any interaction strength. As we will see below, the term $\delta$~(\ref{deltass}) can become comparable with the (conformal) $1/N_a$ correction for 
$N_a\Delta_{\infty}\lesssim 40$ and exceeds higher order power law corrections up to much larger values of $N_a$.

Note that the contribution of the gapped charge excitations for
the repulsive Hubbard model corresponds to the contribution of gapped spin excitations for the  attractive model, and  the contribution of gapless spin excitations corresponds to
the contribution of gapless  charge excitations in the attractive case.

The origin of the term $\delta$ is the following. The state with two additional or two missing spin-up
(or spin-down) particles of the initial attractive Hubbard model contains unpaired fermions with energies above the gap. Gapped excitations of the  model are $S=1/2$ - solitons
which appear only  in pairs.
We  thus calculate the  exact energy of the states with  2 solitons which have  different nonzero momenta and
 energies near the bottom of the excitation band. The contribution (\ref{g0})
takes into account these  nonzero kinetic energies of  the solitons and
 is proportional to the curvature of the excitation spectrum.

It is worth noting that the term $\delta$  is  independent of the definition of the gap, except for  minor changes  for a relatively  small number of particles. For example, it
remains the same
(as well as  all other results of our work) if we consider  the gap in the spectrum of triplet excitations:
$2 \Delta = E(N/2 + 1, N/2 -1, -|U|) - E((N/2, N/2, -|U|) $.  In this case  we just get  two unpaired spin-up fermions.

 The calculation for excited states of the system with unpaired  fermions is
 the same as above. It simply assumes that $k_h$ are no longer  fixed by the condition $Z_c (k_h) = (N_a -1)/2N_a$,
  but are related to the  momenta $\pm p$  of the solitons.
  For  the energy of the state containing two solitons with momenta $p$ and $-p$ we have
 $\eps  = 2\tilde{\Delta} + 2 [\eps_c (\pi -p) - \eps_c (\pi)]$, and for low-lying excitations at sufficiently large number of particles it is reduced to
$\eps=2\tilde{\Delta}+\eps''(\pi)p^2$, where $\tilde{\Delta} = \Delta_{\infty} + \delta \Delta_{ng} + \delta_g$.
The minimum value of $\eps$ is achieved at the minimum possible value of $p$ which is equal to $(\pi-k_h^+)$,
 and we arrive at the  gap $\Delta = \Delta_{\infty} + \delta \Delta$, with
 $\delta \Delta$ given by Eq.(\ref{g00}) and thus  including $\delta$ (\ref{g0}).
It is important that this  $\Delta$ is  just the gap
 that is  measured by the radiospectroscopy method used for obtaining the gap in experiments with two-component 3D Fermi gases \cite{exp}.
\section{Exponential corrections for half filling}

Throughout the paper we discuss the case where the inequality 
\begin{equation}            \label{Nau}
N_a u\gg 1
\end{equation}
is satisfied (except for Section VII), so that all analytical formulas obtained below remain valid for arbitrarily large $N_a$.  It is convinient to present the results in terms of
the parameters $N_a\Delta_{\infty}(u)$ and $u$ because in the limit of $u\ll 1$ the most important contributions to both $\delta\Delta_g/\Delta_{\infty}$ and
$\delta\Delta_{ng}/\Delta_{\infty}$  depend only on $N_a\Delta_{\infty}$ (see below).

In this Section we calculate the finite size corrections to the gap, originating from the gapped sector
and given by Eq. (\ref{Gap_Spin_Sector}).
Using Eqs.    (\ref{zc}), (\ref{20}), (\ref{IJ}) and the Poisson relation
$\sum_m \delta (x -mn) = \sum _m \exp  [ 2 \pi x m i]$ we  obtain
\begin{equation}
\int_{-\pi}^{\pi}  \eps_c(k) X_N^c (k) \; d k =\\
-\int_{-\pi}^{\pi}  \eps_c (k) \left[ \frac{\rho_N (k)}{\exp[-2 \pi i N_a Z_N^c (k + i 0)] + 1} +\frac{\rho_N (k)}{\exp[2 \pi i N_a Z_N^c (k - i 0)] + 1}\right] \; dk.
\label{int}
\end{equation}
To lowest order,  we take the functions $\rho_N (k)$ and $Z_N^c (k)$ equal to their values
 in the thermodynamic limit. So,  they are given by Eq. ~(\ref{Densities_Hubbard_Model_Infinity})
and Eq. (\ref{zn})
for the case $N=N_a$, and      by
  Eq. ~(\ref{Densities_Hubbard_Model_Na-2_Infinity-k})  and Eq. (\ref{zn-2}) for
 $N=N_a-2$, with $k_h=\pm \pi$ and $\lambda_h=\pm\infty$.
One can show that the integral from $-\pi$ to $\pi$ in the right -hand side of Eq. (\ref{int}) is equal to the integral from $- \pi + i \mbox{arcsinh} u$ to $ \pi + i
\mbox{arcsinh} u$ for  the first term of the integrand plus the integral from$- \pi - i \mbox{arcsinh} u$ to $ \pi - i \mbox{arcsinh} u$ for the second term.

Equation (\ref{Gap_Spin_Sector}) then reduces to
\bea
\label{Saddle_Point}
\frac{\delta \Delta_g}{N_a} = \int_{-\pi + i\gamma}^{\pi + i \gamma}\left[\frac{\rho_{N_a} (k) \eps_c (k)}{ 1+
\exp \left[-2 \pi i N_a Z^c_{ N_a} (k)\right]} -  \frac{\rho_{N_a -2 } (k) \eps_c (k)}{1+ \exp \left[-2 \pi i N_a Z^c_{ N_a -2} (k)\right]}\right] \; dk  \nonumber\\
+ c. c. = \frac{1}{2 \pi i N_a}\int_{-\pi + i\gamma}^{\pi + i \gamma}
\eps'_c (k) \ln \left(\frac{
1 +   \exp [2 \pi i N_a Z^c_{ N_a} (k)]}{1 +   \exp [2 \pi i N_a Z^c_{ N_a -2} (k)]}\right) dk + c.c.,
\eea
where  $\gamma = \mbox{arcsinh} u$.

\begin{figure}[t]
\centerline{
	\mbox{\includegraphics[width=3.0in]{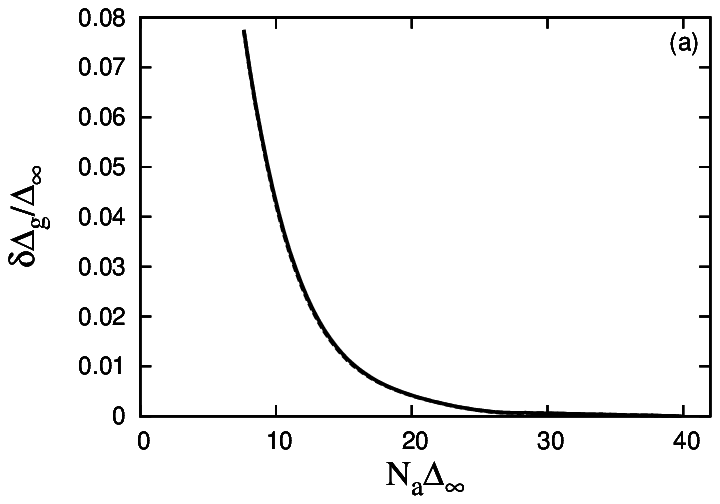}}
}
\centerline{
	\mbox{\includegraphics[width=3.0in]{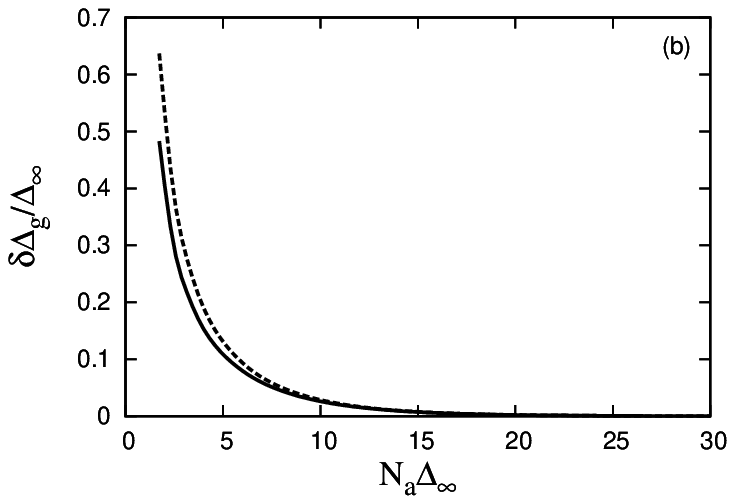}}
}
\centerline{
	\mbox{\includegraphics[width=3.0in]{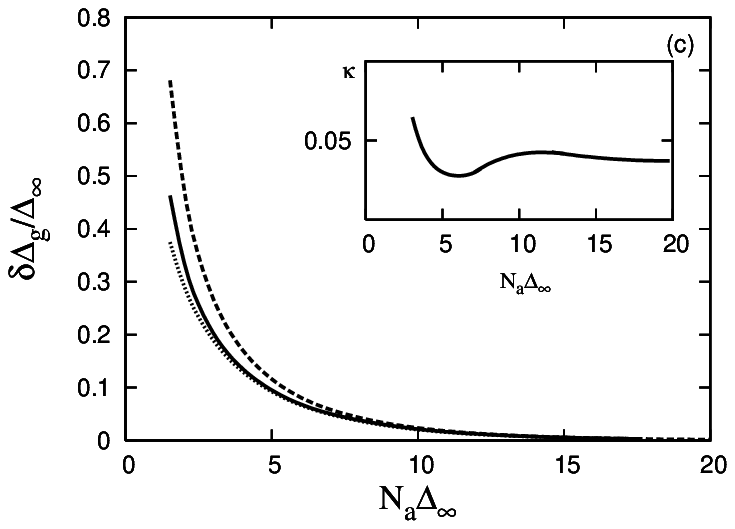}}
}
\caption{The correction $\delta \Delta_g / \Delta_{\infty}$ versus
$N_a \Delta_{\infty}$. The solid curve is the result of exact calculations from Eq. (\ref{Gap_Spin_Sector}),
and  the dashed curve represents the result of Eq. (\ref{Saddle_Point}).
In a) $u = 1$, in b) $u = 0.5$,  and in c) $u  = 0.25$ with the dotted curve showing the  result of Eq. (\ref{ull1}), with the replacement $N_a\rightarrow N_a(1-u/\pi)$ 
(see text after Eq.~(\ref{ull1})). The inset in c) shows the relative difference between the exact $\delta\Delta_g$ and the exponential result of equation (\ref{ull1}),
$\kappa=(\delta\Delta_g^{\rm exact}-\delta\Delta_g^{\rm exp})/\delta\Delta_g^{\rm exact}$.}
\label{Figa}
\end{figure}

It is convenient to present the ratio $\delta \Delta_g/\Delta_{\infty}$ as a function of  $N_a \Delta_{\infty}$
and  $u$.  In Fig. \ref{Figa}  we show  $\delta \Delta_g/\Delta_{\infty}$ versus $N_a \Delta_{\infty}$ for
several values of $u$, and one  clearly sees that for  not very large
$N_a \Delta_{\infty}$ this correction becomes significant.

The edge points of  the integration    $k_0=\pm \pi\pm i \mbox{arcsinh} u$  are
the saddle points at which  $\rho_{N_a}(k)=dZ_{N_a}/dk=0$.
For  sufficiently large $N_a$ we may use  the saddle point approximation, and the  expression for  $\delta \Delta_g$ becomes
\bea
\label{Correction_Charge}
\delta \Delta_g  \approx C
\frac{|\eps'_c (k_0)|
  }{\pi \, \sqrt{N_a|\rho'(k_0)|}}  \exp[- S_0],
\eea
where
\bea
\label{sp2}
\eps'_c(k_0) &=& i \left[2u - 2 \sqrt{u^2 + 1}\int_0^{\infty} J_1 (\om)  \tanh (\om u) \, \exp (-\om u)\; d\om \right], \\
 \rho'(k_0) &=& \frac{i}{2 \pi} \left[\frac{u}{\sqrt{u^2 + 1}} + (u^2 + 1)\int_0^{\infty} \om \tanh (\om u) \; J_0(\om) \exp (-\om u)\; d \om \right], \\
S_0 &=& - 2 \pi i (Z^c_{N_a} (k_0) - 1/2 )=  N_a \left[\gamma - \int_0^{\infty} \frac{\tanh ( \om u) \exp (-\om u)}{\om }J_0(\om) d \om \right],\\
C &=&   \left[1 - \left(\frac{\Gamma (3/4)}{2 \Gamma (5/4)}\right)^4 \right],
\label{At_The_Saddle_Point-last}
\eea
and we used the  relation
\be
Z_{N_a -2}^c (k_0 ) = Z_{N_a}^c (k_0) + \frac{i}{\pi N_a} \int_0^{\infty}\frac{\tanh (\om u)}{\om}
 \exp (- \om u) \;d \om = Z^c_{N_a} (k_0) +\frac{2i}{\pi N_a}
\ln \left[ \frac{2\Gamma(5/4)}{\Gamma(3/4)}\right].
\ee
The saddle point approximation assumes that
the exponent in Eq. (\ref{Correction_Charge}) is large:
\beq
S_0 \gg 1 .
\label{ss>1}
\eeq
In the  case of strong interaction, $u \gg 1$,  Eq. (\ref{Correction_Charge}) gives
\beq
\delta \Delta_g \approx \frac{1}{\sqrt{N_a}\,  u^{N_a -1}},
\label{ugg1}
\eeq
and one sees that in this limit the correction $\delta \Delta_g$ is negligible.

The situation  changes  for $ u < 1$.
In the limit of $u \ll 1$,  from Eq. (\ref{Correction_Charge}) we obtain:
\bea
\label{ull1}
\delta \Delta_g  \approx C \sqrt{ \frac{{2}}{ { \pi}}}
 \frac{ \Delta_{\infty} \exp[ - \Delta_{\infty} N_a /4]}{\sqrt{\Delta_{\infty} N_a}} \approx 0.63 \frac{ \Delta_{\infty} \exp[ - \Delta_{\infty} N_a /4]}{\sqrt{\Delta_{\infty}
N_a}},
\eea
where the gap in the  thermodynamic  limit,  $\Delta_{\infty} $, is given by Eq. (\ref{ddd}).
The criterion
(\ref{ss>1})  then becomes $
N_a \Delta_{\infty} \gg 1$.
The obtained  relation    (\ref{ull1})   is in accordance with
 the  universal scaling behavior of the gap in massive quantum field theories \cite{lusher}, which is expected
 for the Hubbard model at $u \ll 1$. For not very small $u$ we should take into account corrections which are linear in $u$ 
in the expression for $S_0$ and in the preexponential factor for $\Delta_{\infty}$. This proves to be equivalent to the replacement $N_a \to N_a (1 -u/\pi )$ in Eq.~(\ref{ull1}).
Higher order corrections become important only for $N_a\Delta_{\infty}$ which are so large that the exponential contribution $\delta\Delta_g$ is no longer important.

\section{Power law corrections}
\begin{figure}[t]
\includegraphics[width=4.0in]{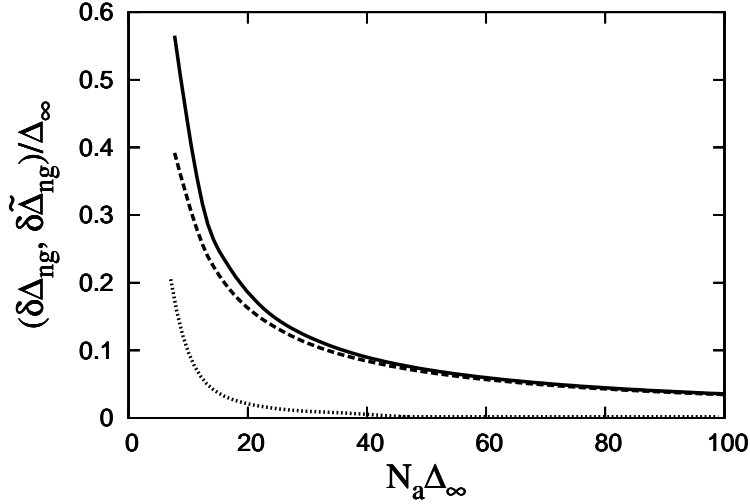}
\caption{The correction $\delta \Delta_{ng} / \Delta_{\infty}$ versus $N_a \Delta_{\infty}$ for $u =1$.
The solid curve shows the result of exact calculation from Eq. (\ref{gss}) and the dashed curve the result of Eq. (\ref{accurate}). The dotted curve represents higher order power
law corrections $\delta\tilde\Delta_{ng}$ (see text).}
\label{Fig}
\end{figure}

The correction to the gap provided by the gapless sector, $\delta \Delta_{ng}$, we calculate using the  conformal field theory $1/N_a$ expansion for the energy
\cite{w1,w3,fr-kor}:
\be
E_{N_a} - E_0 = \frac{2 \pi}{N_a} v_s (\Delta_+ + \Delta_- ),\qquad E_0 = \eps N_a - \frac{\pi}{6 N_a} v_s ,
\ee
where $\eps N_a$ is the  energy  in the thermodynamic limit,  and the velocity of
 spin excitations  is given by
\be
v_s = \frac{d\eps}{d p} = \frac{\eps_s^{\prime} (B)}{2 \pi \sigma (B)},
\ee
where $B$ is defined by Eq. (\ref{Number_Particles_Hubbard_Model}).
The conformal dimensions of primary operators are equal to
\be
\Delta_{\pm} = \frac{1}{2} \left( \xi_s(B) D_{s} \pm
\frac{\Delta N_s}{2 \xi_s (B)} \right)^2.
\ee
Integer or half-odd integer numbers $\Delta N_s$, $D_s$ are characterizing  the excitation states.
The component of the charge dressed matrix $\xi_s$ is determined   from the equation
\be
\xi_s (\lambda) =
1- \frac{1}{2\pi} \int_{-B}^{B} K_2(\lambda -\eta)\xi_s (\eta) d \eta.
\ee
For the zero-field case ($B = \infty$) the solution is $\xi_s  = 1/\sqrt{2}$.
The  energy  and  momentum of a spin excitation are given by
\bea
\eps_s (\lambda ) &=& 2 \int_0^{\infty}\frac{\cos \om \lambda}{\cosh \om u }J_1 (\om)
\frac{d \om }{\om},\\
p_s (\lambda) &=& \frac{\pi}{2} - \int_0^{\infty}\frac{\sin \om \lambda}{\cosh \om u }
J_0 (\om)\frac{d \om }{\om}, \\
\frac{dp_s}{d \lambda} &=& - 2 \pi \sigma(\lambda).
\eea
The excitation state with the energy $E(N_a/2, N_a/2 -2)$ is characterized by the numbers
$D_s =0, \;\; \Delta N_s = 1$ . According to Eq. (\ref{Gap_Hubbard_Model3}), this leads to the
correction
\be
 \delta \Delta_{ng} = \frac{ \pi}{2 N_a} \frac{v_s}{\xi_s^2}.
\ee
 For the half-filled  case  we obtain
\be
 \delta \Delta_{ng} = \frac{2 \pi}{N_a} \frac{I_1 ({\pi}/{2 u})}
{I_0 ({\pi}/{2 u})}.
\label{Correction_Spin}
\ee
A more accurate result following from the calculations in Refs. \cite{w1,fr-kor} reads:
\begin{equation}
\label{accurate}
\delta \Delta_{ng}= \frac{2 \pi}{N_a} \frac{I_1 ({\pi}/{2 u})}{I_0 ({\pi}/{2 u})}\left(1-\frac{1}{2\ln[N_aI_0(\pi/2u)]}\right).
\end{equation}
The comparison of Eq.~(\ref{accurate}) with the result of exact calculations from Eq. (\ref{gss}) shows the validity of Eq.~(\ref{accurate})
even for not very large $N_a$ (see Fig. \ref{Fig}).  For example, at $u = 1$ even for $N_a = 10$
($N_a \Delta_{\infty} \approx 13$ ) the relative difference is $\sim 20\% $. The substraction of $\Delta_{ng}$~(\ref{accurate})
from the exact result of Eq.~(\ref{gss}) gives higher order power law corrections which we denote as $\delta\tilde\Delta_{ng}$. For $u=1$ they are represented by the dotted curve
in Fig.~\ref{Fig}.

In the limit of strong  coupling Eq. (\ref{Correction_Spin}) yields:
\beq
\delta \Delta_{ng}  \backsimeq \frac{\pi^2}{2 N_a u}; \qquad u \gg 1,
\label{ugg11}
\eeq
and the power law correction (\ref{ugg11}) always dominates over the negligible exponential correction (\ref{ugg1}).
The situation changes for $u\apprle 1 $.
For not very large $N_a\Delta_{\infty}$, the exponential
correction $\delta \Delta_g$  originating from gapped
excitations becomes comparable with $\delta \Delta_{ng}$ and exceeds higher order power law corrections. We provide a detailed comparison of $\Delta_g$ with power law corrections
in the next section.

\begin{figure}[h!]
\centerline{\mbox{\includegraphics[width=3.0in]{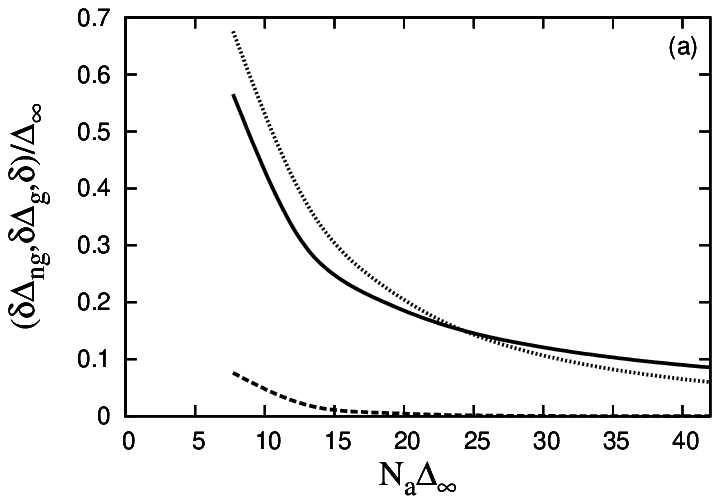}}\mbox{\includegraphics[width=3.0in]{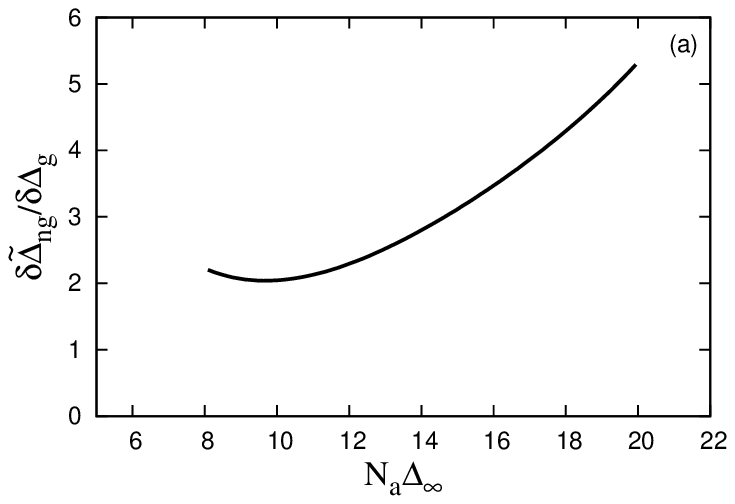}}}
\centerline{\mbox{\includegraphics[width=3.0in]{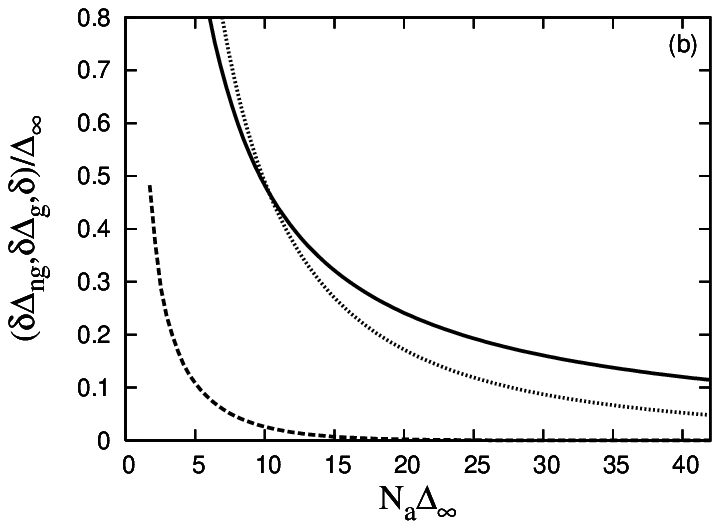}}\mbox{\includegraphics[width=3.0in]{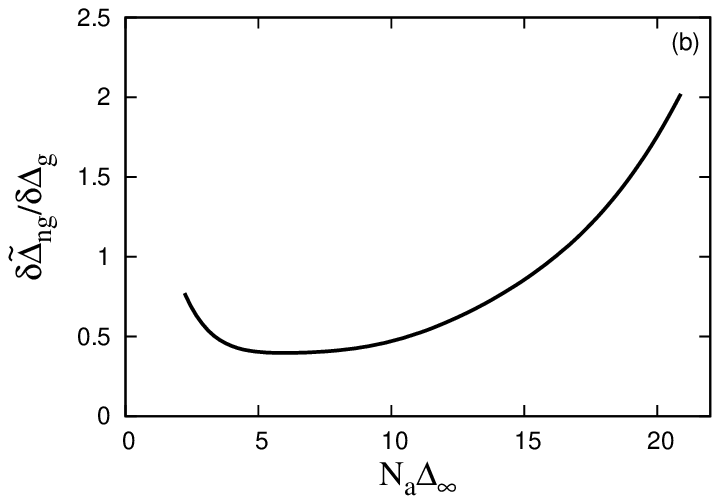}}}
\centerline{\mbox{\includegraphics[width=3.0in]{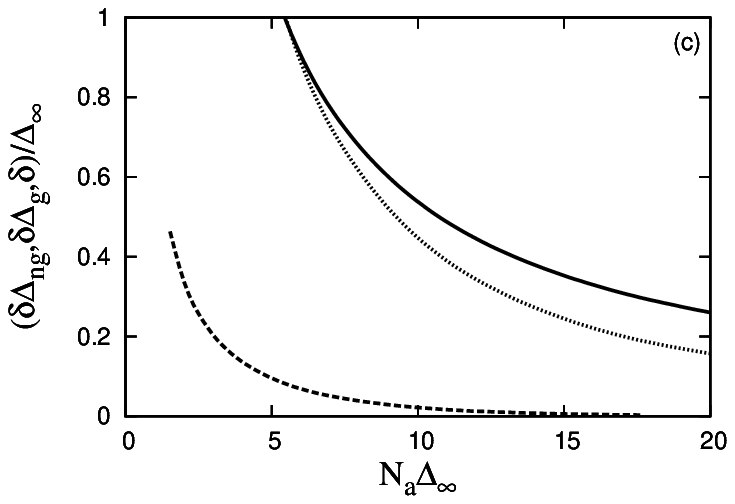}}\mbox{\includegraphics[width=3.0in]{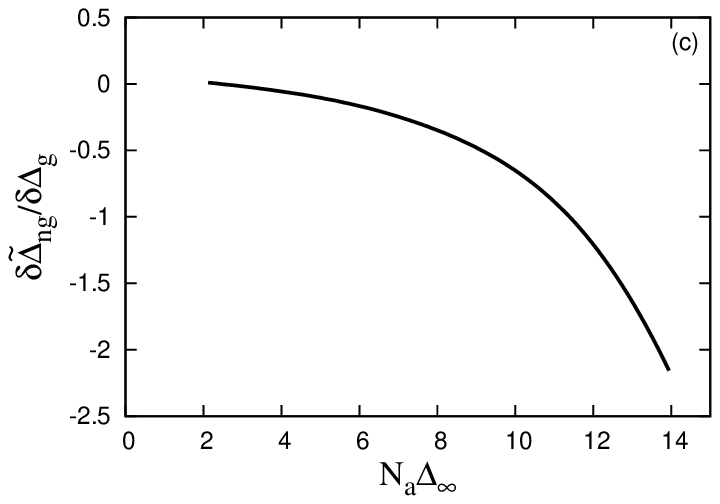}}}
\caption{The result of exact calculations  from Eqs. (\ref{gss}), (\ref{Gap_Spin_Sector}) and (\ref{g0}).
In the left part the solid curve shows $\delta \Delta_{ng} / \Delta_{\infty}$ versus $N_a\Delta_{\infty}$, the dashed curve $\delta \Delta_{g}/\Delta_{\infty}$,
and the dotted curve $\delta/\Delta_{\infty}$. In the right part the solid curve shows the ratio of the higher order power law corrections $\delta\tilde\Delta_{ng}$
to the exponential correction $\delta \Delta_{g}$. In a) $u = 1$, in b) $u = 0.5$, and in c) $u = 0.25$.}
\label{Fig5}
\end{figure} 

\section{Discussion of the results}

We start the comparson of the non-conformal exponential correction to the gap with power law corrections in the limiting case of $u\ll 1$ and $N_a\Delta_{\infty}\gg 1$,
so that Eq.~(\ref{ull1}) is applicable. Then, comparing the result of Eq.~(\ref{ull1}) with that of Eq.~(\ref{accurate}) we see that the power law $1/N_a$ correction dominates
over the exponential correction for any $N_a\Delta_{\infty}$ significantly larger than unity. However, the exponential corection (\ref{ull1}) is still larger than higher order
power law corections
in a wide range of $N_a$. In the limit of $N_a\rightarrow\infty$ the higher order corrections contain terms $\ln[\ln(N_a)]/(N_a\ln^q(N_a))$,  $1/(N_a\ln^q(N_a)$, $1/N_a^2$, etc.,
where $q\geq 2$ is an integer \cite{w1}. Thus, for reasonable values of $N_a$ the term that should be compared with $\delta\Delta_g$ (\ref{ull1}) is
$\sim 1/[N_a\ln^2(N_a)]$. The argument of the logarithm may contain an $u$-dependent multiple $B\sim \exp(\pi/2u)$ like the logarithm in the second term of Eq.~(\ref{accurate}).
This is however not important. According to Eq.~(\ref{ddd}), in the limit of $u\ll 1$ the gap is exponentially small, $\Delta_{\infty}\sim\exp(-\pi/2u)\ll 1$. Hence, for
$N_a$ at which the exponential corection can still be important we have $N_a\Delta_{\infty}\ll\Delta^{-1}_{\infty}$ and $\ln(N_a)\approx\ln(\Delta^{-1}_{\infty})\sim 1/u$. So, 
irrespective of the presence of the $u$-dependent multiple $B$, the higher order power law corection becomes $\sim (u^2/N_a)\ln(1/u)$ and it is smaller than the exponential
correction 
(\ref{ull1}) for $N_a$ satisfying the inequality
\begin{equation}         \label{maincond}
N_a\Delta_{\infty}\lesssim 8\ln\left(\frac{1}{u}\right);\,\,\,\,\,\,\,\,\,u\ll 1.   
\end{equation}
Thus, at $u\ll 1$ for $N_a\Delta_{\infty}$ significantly larger than unity but still satisfying the condition (\ref{maincond}), the exponential correction (\ref{ull1}) is
legitimate and can be kept together with the power law correction (\ref{accurate}).

The situation is similar for intermediate values of $u$ smaller than unity. This is seen from Fig.~3 where we present our numerical results or $u=0.25$ and $u=0.5$. However,
already for $u=1$ the exponential correction is smaller than the higher order power law terms (see Fig.~3) and should be omitted. As far as the term $\delta$ is concerned, for
$u\ll1$ it is comparable with the power law correction up to $N_a\Delta_{\infty}\sim 40$ and exceeds higher order power law corrections for much larger $N_a$. Even if we compare
$\delta$ with the second term of Eq.~(\ref{accurate}), the latter is smaller at $N_a\Delta_{\infty}\lesssim 1/u$.

\begin{figure}[h!]
\centerline{
	\mbox{\includegraphics[width=3.2in]{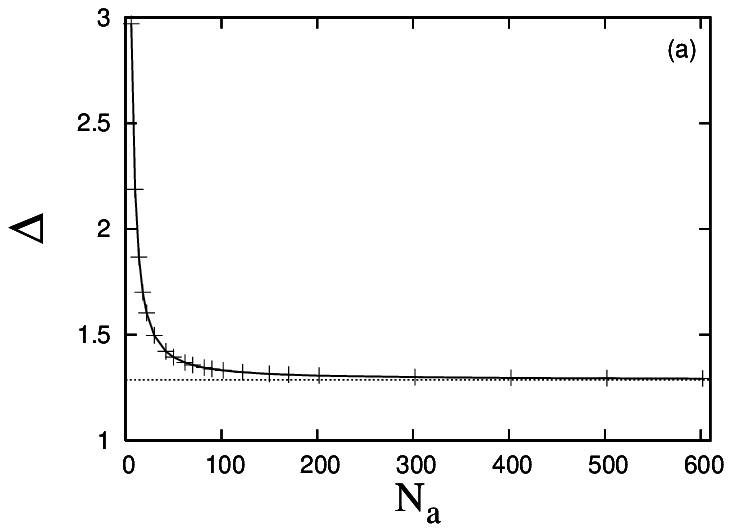}}
}
\centerline{
	\mbox{\includegraphics[width=3.2in]{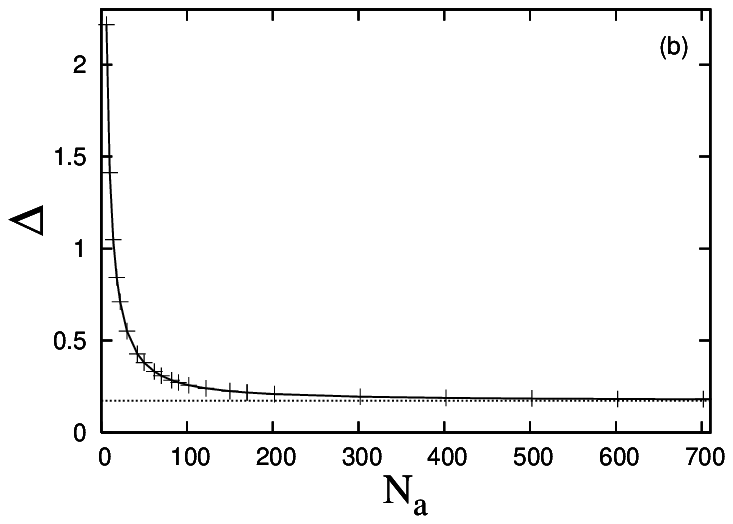}}
}
\centerline{
	\mbox{\includegraphics[width=3.2in]{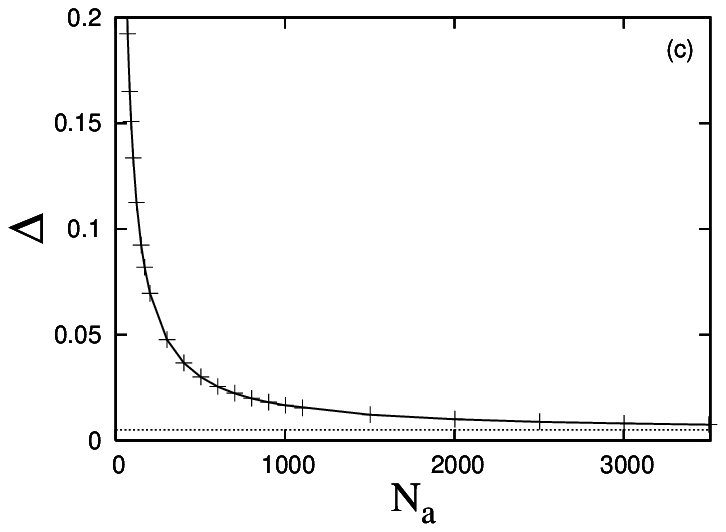}}
}
\caption{The gap $\Delta$ in units of $t$ versus  $N_a$ at half filling for  $u=1$ in a), $u =0.5$ in b),
and $u =  0.25$ in c). The dotted line is the value of the gap in the thermodynamic limit. The sum
$\delta\Delta_{ng}+\delta\Delta_g+\delta+\Delta_{\infty}$ (solid curve) coincides with the value
of the gap found directly (crosses) from the Bethe Ansatz equations (\ref{Discrete_Bethe_Ansatz_Hubbard_Model})
and (\ref{dba2}) using Eqs.~(\ref{energy-discrete}) and (\ref{Gap_Hubbard_Model3}).
}
\label{Fig1}
\end{figure}

In Fig. \ref{Fig1} we present the results of exact calculations  for the gap $\Delta$ at half filling from
Eqs. (\ref{Discrete_Bethe_Ansatz_Hubbard_Model}), (\ref{dba2}), (\ref{energy-discrete}) and
(\ref{Gap_Hubbard_Model3}) for the same $u$ as in Fig. \ref{Fig5}. A direct comparison of $\Delta$ with the gap of the thermodynamic linit $\Delta_{\infty}$  (\ref{h0}) shows that
finite size corrections
can be  safely omitted for $N_a \Delta_{\infty} > 40$. On the other hand, already at $N_a \Delta_{\infty} < 10$, they dominate the gap.
 For $u = 0.25$ where
 $\Delta_{\infty} = 5\cdot 10^{-3}$ this occurs already at $N_a < 1.5 \cdot 10^3$. For $u = 1$
we have $\Delta_{\infty} = 1.28$ and finite size corrections are important only for $N_a < 30$.

It is important that for $N_a \Delta_{\infty} \sim 10$ or even somewhat larger the non-conformal
corrections originating from the gapped sector become comparable with power law corrections coming from the gapless sector. This is seen in
Fig. \ref{Fig5}. The exponential correction $\delta \Delta_g$ is about $20\%$ of $\delta \Delta_{ng}$ or smaller, and the correction $\delta$ 
approaches $\delta \Delta_{ng}$ and can even exceed it.

\begin{figure}[h!]
\centerline{
	\mbox{\includegraphics[width=3.0in]{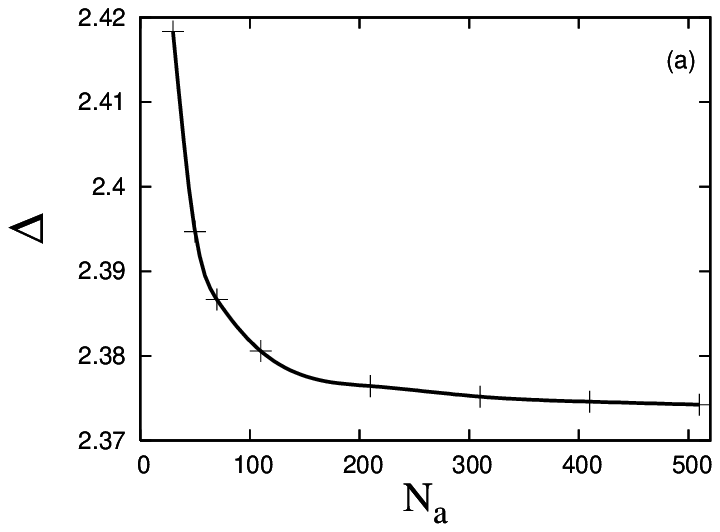}}
}
\centerline{
	\mbox{\includegraphics[width=3.0in]{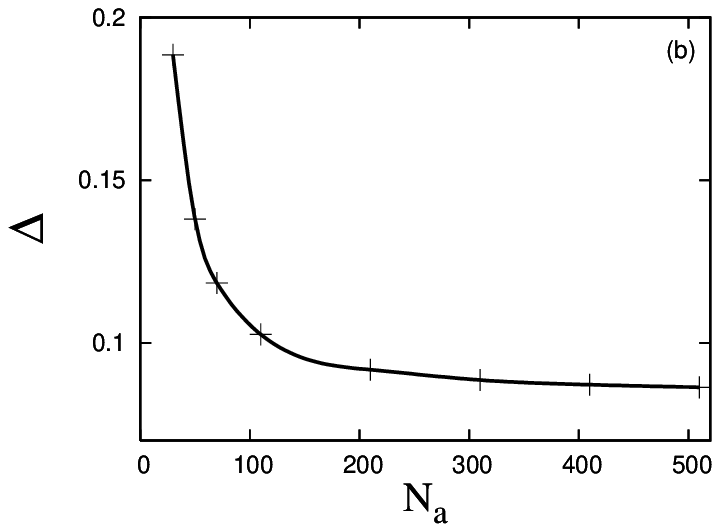}}
}
  \caption{The gap $\Delta$ in units of $t$ versus $N_a$, calculated numerically  for $u=0.25$ in a),
and $u=1.25$ in b)
for the filling factor $n=0.2$.}
\label{Fig3}
\end{figure}

Qualitatively, the dependence  $\Delta (N_a)$ remains the same for smaller filling factors. This is  seen from Fig. \ref{Fig3}, where  we present  our numerical results for $n =
0.2$. For $u = 1$ finite size effects become important
only at a very small number of lattice sites $N_a < 30$. For $u = 0.25$ the
thermodynamic-limit gap is $\Delta_{\infty} \approx 6 \cdot 10^{-2}$ and finite size effects are already important for $N_a \approx 500$.

\section{Limit of $N_a u\ll1$}
In the limit of  $N_a u\ll1$, which can be realized for $u \ll 1$,
the energy
spectrum of the attractive Hubbard Model shows no exponential gap and both charge and spin sectors are conformal.
The analysis of Lieb-Wu equations for this case  has been done in \cite{fr,am}, and found corrections
 $\sim u/N_a$. This can be understood from the conformal $1/N_a $ expansion as a consequence of the
linear  dependence of the velocities of elementary excitations
on the interaction constant $u$ \cite{fr,chub}.

Here we  consider the limit of $u N_a \ll 1$  for completeness and
present first order  corrections in  $u$ to the ground state energy and  to the gap in  the excitation spectrum.
 As  in the previous sections, we  calculate the  energy for the Hubbard model in the repulsive case,
 where  the Bethe Ansatz equations are easily solved, and then
 restore  the energy for the attractive case using the particle-hole symmetry.

So, consider   a system of $N$ particles ($N_{\downarrow}$  spin-down and  $N - N_{\downarrow}$
spin-up)  with repulsive interaction.
From the Lieb-Wu equation (\ref{Discrete_Bethe_Ansatz_Hubbard_Model}) we obtain
 the momenta $k_j$ to  first order in $u$:
\begin{equation}
\exp ({ik_jN_a})=\prod _{\alpha=1}^{N_{\downarrow}}
\frac{\sin  k_j-\lambda_{\alpha}+iu}{\sin  k_j-\lambda_{\alpha}-iu}\Rightarrow
\delta k_j=k_j - k_j^0 =\frac{1}{N_a}\sum_{\alpha=1}^{N_{\downarrow}}
 \frac{2u}{\sin  ( k_j^0 )-\lambda_{\alpha}^0},
\end{equation}
where $k^0$ and $\lambda^0$ are the momenta and rapidities for $u \to 0$.
The energy itself and the  interaction-induced change of the energy are given by
\begin{equation}
E=-2\sum_{j=1}^{N} \cos  k_j , \quad
\delta E=E-E^0=\frac{4u}{N_a}\sum_{j=1}^{N}\sum_{\alpha=1}^{N_{\downarrow}}
\frac{\sin k_j^0}{\sin (k_j^0)-\lambda_{\alpha}^0},
\end{equation}
where $E_0$ is the ground state energy for $u \to 0$.
We  now  calculate  the densities of momenta $k$ and rapidities $\lambda$ in the thermodynamic limit
from Eqs.~(\ref{Thermodynamic_Bethe_Ansatz_Hubbard_Model-k}) and (\ref{Thermodynamic_Bethe_Ansatz_Hubbard_Model-lambda}).
To the lowest order in $u$ we obtain:
\bea
\label{76-1}
2\sigma(\lambda)&=&\int_{-Q}^{Q}\delta(\lambda-\sin k)\rho(k)dk,\\
\rho(k)&=&\frac{1}{2\pi}+\cos k\int_{-B}^{B}\delta(\lambda-\sin k)\sigma(\lambda)d\lambda.
\label{76-2}
\eea
The solution of Eqs. (\ref{76-1}) and (\ref{76-2}) is ($ n_{\downarrow} \leq n_{\uparrow}$)
\begin{eqnarray}
\label{Densities_Small_NaU-k}
\rho(k)&=&\left\{\begin{array}{ll}
				1/\pi; \quad & { k \leq \pi n_{\downarrow}},\\
        1/2\pi; \quad & { \pi n_{\downarrow}< k \leq \pi(n-
        n_{\downarrow})}.\end{array} \right.\\
\label{Densities_Small_NaU-lambda}
\sigma(\lambda)&=&\frac{1}{2\pi}\frac{1}{\sqrt{1-\lambda^2}}.
\end{eqnarray}
Then, using Eq.~(\ref{Number_Particles_Hubbard_Model}) for the total number of particles and the number of spin-down particles, we find an expression for the integration limits $Q$
and $B$:
\be
B=\sin\left(\pi n_{\downarrow}\right),\qquad Q=\pi(n-n_{\downarrow}).
\ee
Using  Eqs.~(\ref{Densities_Small_NaU-k}) and (\ref{Densities_Small_NaU-lambda}) we obtain the interaction-induced change of the energy to first order
in $u$:
\begin{eqnarray}
\frac{\delta E}{N_a}&=&4u\int_{-Q}^{Q}  d k \int_{-B}^{B} \frac{\sin k}{\sin k-\lambda}\sigma(\lambda)\rho(k) \; d \lambda =
\nonumber\\
&=&\frac{4u}{\pi^2}(\pi  n_{\downarrow})^2+\frac{4u}{\pi^2}\int_{\pi  n_{\downarrow}}^{\pi( n- n_{\downarrow})}dk\int_{0}^{\pi  n_{\downarrow}} \frac{\sin^2 k}{\sin^2 k-\sin^2 q} d
q,
\label{79}
\end{eqnarray}
where we turned to the variable   $q = \arcsin\lambda$.
For the case of attraction we should substitute $n_{\uparrow} \to 1 - n_{\uparrow}$, $n_{\downarrow} \to  n_{\downarrow}$, in accordance with the symmetry properties
(\ref{Symmetries_Hubbard_Model}). Integrating over $dq$ we find:
\beq         \label{80}
\frac{\delta E (-u)}{N_a} = - 4 u n_{\downarrow} + 4 u n^2_{\downarrow} +\frac{2 u}{\pi^2}
\left( 2 \int_{\pi n_{\downarrow}}^{\pi /2} - \int_{\pi n_{\downarrow}}^{\pi n_\uparrow} \right)
  \tan k  \; \ln \left( \frac{\tan k + \tan \pi n_{\downarrow}}{\tan k - \tan \pi n_{\downarrow}}\right) dk,
\eeq
where $n_{\downarrow}$ and $n_{\uparrow}$ are already occupation numbers for the attractive Hubbard model, and we  assume that $n_{\downarrow} \leq n_{\uparrow}$.
Eq.~(\ref{80}) leads to the following result for the interaction-induced change of the energy
to first order in $N_a u$:
\begin{eqnarray}
&&\frac{\delta E(-u)}{N_a}=4 u  n_{\downarrow}^2- 4 u n_{\downarrow} + \frac{4 u}{\pi}\arctan \pi n_{\downarrow} -\frac{u}{\pi^2}\ln \left( \frac{1+\tan^2\pi n_{\uparrow}}
{1+\tan^2\pi n_{\downarrow} } \right) \ln \left( \frac{\tan\pi n_{\uparrow}+ \tan\pi n_{\downarrow} }{\tan\pi n_{\uparrow}-\tan\pi n_{\downarrow} } \right)  \nonumber \\
&&+\frac{2u}{\pi^2} \left( -\textrm{Re} Li_2 \left(\frac{2 \tan \pi n_{\downarrow}}{\tan \pi n_{\downarrow} - i} \right) - \textrm{Re} Li_2\left(\frac{\tan \pi n_{\uparrow} +\tan
\pi n_{\downarrow}}{\tan \pi n_{\downarrow} - i}\right) + \textrm{Re} Li_2 \left(\frac{\tan \pi n_{\uparrow} -\tan \pi n_{\downarrow}}{ -\tan \pi n_{\downarrow} -
i}\right)\right), \nonumber
\end{eqnarray}
where $ Li_2\left(z\right)=\sum_{k=1}^{\infty} {z^k}/{k^2}$ is a polylogarithmic function.

In  the  limit of small filling factors, $N_{\uparrow}\ll N_a$ and $N_{\downarrow}\ll N_a$, after a straightforward algebra we obtain:
\begin{equation}   \label{smalln}
\frac{\delta E}{N_a} \approx u(-4n_{\downarrow}  -
4 n_{\uparrow}n_{\downarrow}+ 2 n_{\downarrow}^2)
\end{equation}
The limit of small filling factors in the Hubbard model corresponds to the gas phase of spin-1/2 fermions.  For this case the ground state energy at $Nu\ll 1$ has been calculated
in Refs. \cite{as,fu,Batchelor}, and the result of Eq.~(\ref{smalln}) coincides with that of  Refs. \cite{as,fu,Batchelor} in the attractive case.

Using Eq.~(\ref{2delta}) we then find a small interaction-induced  correction to the gap in the excitation spectrum
($n_{\uparrow} =  n_{\downarrow} = n/2$) of the attractive model to the lowest order in $uN_a$.
For small filling factors we have: $ \delta \Delta \approx  {4 u}/{N_a}$, and
in the considered limit of $N_a u  \ll1$ this correction is small compared to the level spacing $ \sim 1/N_a$ in our finite size system.

\section{Conclusions}

In conclusion, we have studied finite size effects for the gap in the excitation spectrum of the 1D Fermi Hubbard model with one-site attraction. For the
situation in which the thermodynamic-limit gap $\Delta_{\infty}$   exceeds the level spacing (near  the Fermi energy) of the finite size  system, there are  two types of finite
size corrections. For large interactions ($u \gg  1$) the leading is a power law conformal correction to $\Delta_{\infty}$, which behaves as $1/N_a$ and originates from the gapless 
sector of the excitation spectrum. We also find  non-conformal  corrections originating from the gapped branch of the spectrum. As found at half filling,
in the weakly interacting regime ($u \apprle  1$) the non-conformal corrections
can become of the order of the conformal correction even for the number of particles (lattice sites) as
large as $\sim 20/\Delta_{\infty}$. 
Also, for $u\ll 1$ and large $N_a\Delta_{\infty}$, the exponential correction (\ref{ull1}) is legitimate as long as the condition (\ref{maincond})
is satisfied. Thus, we have the full right to take it into acount together with the power law corection (\ref{accurate}).

For sufficiently small number  of lattice sites (particles) the gap $\Delta$ is
dominated by finite size effects. From a general point of view, this  happens when $\Delta_{\infty} \apprle 1/N_a$, i. e.  $\Delta_{\infty}$ is smaller than the  level spacing of
the finite size system at energies close to the Fermi
energy.  Accordingly, for large interactions ($u \gg 1$) the finite size effects
 are not  important as long as $N_a\gg 1$. However, in the  weakly interacting
regime ($u\apprle 1$)
 they  become  dominant already at significantly
larger $N_a$  than a simple dimensional estimate $1/\Delta_{\infty}$. This is  clearly seen from our results in Fig. 3 and Fig. 4 for $\Delta (N_a )$  at
half filling.

Our findings are especially  important for the studies of the  1D regime with cold atoms, where the number of particles in  a 1D  tube ranges from several tens to  several hundreds
\cite{xxx,xxx1}.  For such a system in the weakly interacting  regime one can not  use  the result of the  thermodynamic limit for the gap. Consequently, one can not employ  the
local density approximation for $\Delta$ based on this result, for  finding the spectrum of isospin gapped excitations in an external harmonic potential.

\section*{Acknowledgements}
We are grateful to A. M. Tsvelik for fruitful discussions and acknowledge hospitality and support of Institut Henri Poincar\'e during the
workshop ``Quantum Gases'' where part of this work has been done.  We also aknowledge D.L. Kovrizhin for his useful suggestions on numerical methods.
The work was supported by the IFRAF Institute, by
ANR (grants 05-BLAN-0205 and 06-NANO-014-01), by the QUDEDIS program of ESF, and by the Dutch Foundation FOM.
LPTMS is a mixed research unit No. 8626 of CNRS and Universit\'e Paris Sud.


\end{document}